\documentclass[12pt]{article}
\usepackage{hyperref,english}

\makeatletter
\def\vereq#1#2{\lower3pt\vbox{\baselineskip1.5pt \lineskip1.5pt
\ialign{$\m@th#1\hfill##\hfil$\crcr#2\crcr\sim\crcr}}}
\makeatother

\begin{document}

\begin{titlepage}
\begin{center}
\hfill LBNL-47184 \\
\hfill UCB-PTH-00/41 \\
\hfill EFI-2000-50 \\
\hfill ANL-HEP-PR-00-135 \\

\vskip 0.2in

{\large \bf Viable Ultraviolet-Insensitive\\
  Supersymmetry Breaking\footnotemark}\footnotetext{This work was
  supported in part by the U.S. Department of Energy under Contracts
  DE-AC03-76SF00098, DE-FG02-90ER40560 and W-31-109-ENG-38, and in
  part by the National Science Foundation under grant PHY-95-14797.  }

\vskip 0.3in

Nima~Arkani-Hamed,$^{1,2}$ David~E.~Kaplan,$^{3,4}$ \\
Hitoshi~Murayama,$^{1,2}$ and
Yasunori~Nomura$^{1,2}$\footnotemark\footnotetext{Research fellow,
  Miller Institute for Basic Research in Science.} 

\vskip 0.1in

{\em $^{1}$Theoretical Physics Group\\
     Ernest Orlando Lawrence Berkeley National Laboratory\\
     University of California, Berkeley, California 94720}

\vskip 0.05in

{\em $^{2}$Center for Theoretical Physics, Department of Physics\\
     University of California, Berkeley, California 94720}

\vskip 0.05in

{\em $^{3}$Enrico Fermi Institute, University of Chicago\\
    5640 Ellis Avenue, Chicago, IL 60637}

\vskip 0.05in
{\em $^{4}$High Energy Physics Division, Argonne National Laboratory\\
    9700 S.Cass Ave, Argonne, IL 60439}

\end{center}

\vskip .1in

\begin{abstract}
  It is known that one can add $D$-term contributions for $U(1)_Y$ and
  $U(1)_{B-L}$ to the anomaly-mediated supersymmetry breaking to make
  the superparticle spectrum phenomenologically viable.  We point out
  that this can be done without spoiling its important virtue, namely
  the ultraviolet insensitivity.  This framework can be derived from
  supersymmetry breaking and $U(1)_{B-L}$ breaking on hidden brane(s).
\end{abstract}

\end{titlepage}

\section{Introduction}

Supersymmetry is one of the most motivated candidates for physics
beyond the standard model because it makes the hierarchy between the
electroweak scale and a fundamental scale, such as the Planck scale,
stable against radiative corrections.  Of course, supersymmetry needs
to be softly broken in Nature as we have not yet observed any
superparticle.  The main challenge in supersymmetric model
building is probably the flavor problem: arbitrary soft breaking of
supersymmetry would induce too-large flavor-changing neutral-current
effects.  This problem can be swept under the rug by making specific
assumptions about the superparticle mass spectrum at the electroweak
scale.  However, justifying such assumptions in a consistent framework
which includes the explanation for the fermion mass hierarchy proves
difficult.  The main reason for the difficulty is that any models
which explain the fermion mass hierarchy must distinguish three
generations of quarks and leptons.  Their structures should also be
sufficiently complicated to generate the observed fermion mass
hierarchy.  For instance, in the so-called Froggatt--Nielsen mechanism
\cite{Froggatt:1979nt}, there are a bunch of heavy fields which
distinguish different flavors and generate the effective Yukawa
couplings when they are integrated out.  These dynamics generically
make the masses of squarks and sleptons non-degenerate, and hence
cause dangerous flavor-changing effects \cite{Kostelecky,KMY}.

One way to avoid such a problem is to assume that superparticle masses 
are generated at an energy scale much lower than the energy scale of 
flavor physics which explains the fermion mass hierarchy.  This is 
indeed the idea behind the gauge mediation of supersymmetry breaking 
\cite{Alvarez-Gaume:1982wy,Dine:1982zb} (and recently revived by 
\cite{DNNS}).  No matter how complicated the flavor physics is, its 
only low-energy consequences are the Yukawa matrices of quarks and 
leptons below the scale of flavor physics.  The superparticle masses 
are generated at much lower energy scales where the only flavor-dependent 
parameters in the Lagrangian are the Yukawa matrices.  This makes the 
flavor physics completely hidden from the superparticle masses and 
hence safe phenomenologically.
The same is true of recently proposed gaugino mediated supersymmetry 
breaking \cite{Kaplan:2000ac, Chacko:2000mi}.  The scale of flavor 
physics should be above the compactification scale where the 
superparticle masses are generated, and there is barely room for 
flavor physics above the compactification scale. 
However, flavor-mechanisms of a purely extra-dimensional nature 
\cite{shining, Arkani-Hamed:2000dc} should work well 
in this context \cite{Kaplan:2000av}.

Anomaly mediation of supersymmetry breaking 
\cite{Randall:1999uk,Giudice:1998xp} is another possibility.  If 
soft supersymmetry-breaking parameters are generated from 
superconformal anomalies, their pattern at an energy scale is 
determined only by the physics at that energy scale.  The only source 
of supersymmetry breaking in this scheme is in the auxiliary component 
of the supergravity multiplet, and hence there is only one free 
parameter: the overall scale of supersymmetry breaking.  Then the 
pattern of supersymmetry breaking does not depend at all on physics at 
higher energy scales.  This is the extremely interesting property: a 
complete ultraviolet (UV) insensitivity.  
Even though the flavor physics occurs at the energy scale below 
the scale where supersymmetry-breaking parameters are generated, 
the resulting low-energy parameters do not remember 
the flavor physics at all.  This happens in a rather 
non-trivial fashion.  Above the scale of flavor physics, there are 
additional interactions which distinguish different generations, and 
the soft supersymmetry-breaking parameters are also highly 
flavor-dependent.  But when heavy particles, such as Froggatt--Nielsen 
fields, are integrated out, the threshold corrections due to the loops 
of heavy particles precisely cancel the flavor-dependence of the 
supersymmetry-breaking parameters.

Despite the appeal of UV insensitivity, anomaly mediation of 
supersymmetry breaking is excluded because of its high predictivity: 
sleptons masses-squared are negative and hence the theory would break 
electromagnetism.  There have been proposed various fixes to this 
problem \cite{Pomarol:1999ie,other,Jack:2000cd}.  
Perhaps the most elegant idea is to use 
non-decoupling effects \cite{Pomarol:1999ie}: 
if a heavy threshold arises due to supersymmetry-breaking 
effects, integrating out the heavy particles would modify the low-energy 
breaking parameters off the anomaly-mediated trajectory.  
Unfortunately, this casts doubt on the very virtue of the UV 
insensitivity; if the non-decoupling physics is flavor-dependent, the 
absence of flavor-changing effects cannot be guaranteed.  

In this paper, we point out that there is a different way to modify 
the anomaly-mediation of supersymmetry breaking in such that the 
complete UV insensitivity is preserved.  In addition, the scalar masses 
can be modified to make slepton masses positive.  This is done by 
introducing more sources of supersymmetry breaking, namely the 
auxiliary components in the abelian gauge multiplets.  The gauge 
multiplets are not dynamical; their only raison d'\^etre is to break 
supersymmetry in their $D$-components.  If the associated fictitious gauge 
symmetry is non-anomalous under the physical gauge groups, the 
$D$-components renormalize among themselves and their contributions to 
the scalar masses are completely determined by the fictitious gauge 
charges of the matter fields.  Because in a given theory there are 
only a finite number of fictitious anomaly-free abelian gauge 
symmetries one can consider, the number of parameters is still 
small.  There are only two additional free parameters in the case of 
the Minimal Supersymmetric Standard Model (MSSM) or Next-to-MSSM (NMSSM): 
$D$-components of $U(1)_{Y}$ and $U(1)_{B-L}$.

The model we present is distinct from that of \cite{Jack:2000cd} 
in the following way.  The authors of \cite{Jack:2000cd} consider 
adding $\sim ({\rm TeV})^2$ Fayet-Iliopoulos terms to the theory to make 
slepton masses-squared positive.  This suggests the existence of an 
additional $U(1)$ gauge symmetry at the weak scale (along with 
standard-model singlets to cancel anomalies).  In our approach, only 
the MSSM or NMSSM fields appear at the weak scale while the weak-scale 
$D$-terms can be generated by breaking $U(1)_{B-L}$ at an arbitrarily 
high scales.  In addition, there is no discussion in \cite{Jack:2000cd} 
of the origin of the Fayet-Iliopoulos terms, nor of the complete UV 
insensitivity of introducing supersymmetry breaking through background 
$D$-terms.

This paper is organized as follows.  In Section~\ref{sec:AM}, we
review the basics of anomaly mediation.  We then show that adding
$D$-terms for non-anomalous abelian symmetries to the anomaly
mediation still gives a consistent trajectory for renormalization-group
equations (RGEs) in Section~\ref{sec:D}.  In particular, we show that 
the pattern of soft parameters is determined completely by physics at 
the energy scale of interest (UV insensitive) with a small number of
independent parameters.  In Section~\ref{sec:model}, we present a
model where one obtains anomaly-mediated supersymmetry breaking
together with $D$-term contributions, by breaking supersymmetry and
the abelian gauge groups on hidden brane(s).  We derive the
superparticle spectrum in Section~\ref{sec:NMSSM}.  In
Section~\ref{sec:neutrino} we show that small Dirac neutrino masses
are generated naturally within our model with order of magnitudes
remarkably consistent with the atmospheric neutrino data.  Finally we
conclude in Section~\ref{sec:conclusion}.

\section{Anomaly Mediation\label{sec:AM}}

Anomaly mediation of supersymmetry breaking assumes that the only source 
of supersymmetry breaking is in the $F$-component of the Weyl compensator 
field $\phi$ in supergravity: $\phi = 1 + \theta^{2} m_{3/2}$.  Here 
and below, we follow the discussions in \cite{Pomarol:1999ie}.  
Because the Weyl compensator appears in front of every dimensionful 
parameter, it appears in particular in front of the UV cutoff of the 
theory, or equivalently, of the renormalization scale $\mu$.  
Therefore, the coefficient of the matter kinetic term
\begin{equation}
        \int d^{4}\theta {\cal Z}_{i}(\mu) Q_{i}^{\dagger} Q_{i},
\end{equation}
where $Q$ denotes a matter chiral multiplet in general, is given by
\begin{equation}
        {\cal Z}_{i}(\mu) = Z_{i} \left( 
        \frac{\mu}{\sqrt{\phi\phi^{\dagger}}} \right).
        \label{eq:Z1}
\end{equation}
Here, $Z_{i}$ is the wave function renormalization factor in the {\it 
supersymmetric}\/ theory.  In general, the expansion of the matter 
kinetic term gives the trilinear couplings and the soft scalar 
masses-squared
\begin{equation}
        \ln {\cal Z}_{i}(\mu) = \ln Z_{i}(\mu)
                + (\theta^{2} A_{i}(\mu) + {\rm h.c.})
                - \theta^{2} \bar{\theta}^{2} m_{i}^{2}(\mu).
        \label{eq:Z2}
\end{equation}
The trilinear couplings for the superpotential term $\lambda Q_{i} Q_{j} 
Q_{k}$ is given by $\lambda (A_{i}+A_{j}+A_{k}) \tilde{Q}_{i} 
\tilde{Q}_{j} \tilde{Q}_{k}$.  Comparing Eqs.~(\ref{eq:Z1}, \ref{eq:Z2}), 
we find
\begin{eqnarray}
        A_{ijk} & = & - \frac{1}{2} 
        (\gamma_{i}+\gamma_{j}+\gamma_{k}) m_{3/2},
        \label{eq:A}  \\
        m_{i}^{2} & = & - \frac{\dot{\gamma}_{i}}{4} |m_{3/2}|^{2}.
        \label{eq:m}
\end{eqnarray}
Here, $\gamma_{i}$ is the anomalous dimension and $\dot{\gamma}_{i} = 
d \gamma_{i}/d \ln \mu$. 

The gauge kinetic term
\begin{equation}
        \int d^{2}\theta\, S(\mu) W^{\alpha} W_{\alpha},
\end{equation}
is renormalized only at the one-loop level due to holomorphy
\cite{holomorphy,AM}, while the physical (canonical) coupling
corresponds to the real superfield $R(\mu)$ with
\begin{equation}
        R(\mu) - \frac{T_{G}}{8\pi^{2}} \ln R(\mu)
                = S(\mu) + S^{\dagger}(\mu) 
                - \frac{1}{8\pi^{2}} \sum_{i} T_{i} \ln {\cal Z}_{i}(\mu),
\end{equation}
where
\begin{equation}
        R(\mu) = \frac{1}{g^{2}(\mu)} 
        + \left( \theta^{2} \frac{2 m_{\lambda}(\mu)}{g^{2}(\mu)} 
        + {\rm h.c.} \right)
        + \theta^{2} \bar{\theta}^{2} R_{D}.
\end{equation}
We are not interested in $R_{D}$ in this paper.  Again using the fact 
that the renormalization scale comes together with the Weyl 
compensator
\begin{equation}
        R(\mu) = g^{-2} \left( \frac{\mu}{\sqrt{\phi\phi^{\dagger}}} \right),
\end{equation}
we find
\begin{equation}
        m_{\lambda} = \frac{\beta(g^{2})}{2g^{2}} m_{3/2}.
        \label{eq:gaugino}
\end{equation}

The one unfortunate feature of the above spectrum of soft parameters is
the sign of the sleptons squared masses.  This comes from the fact that,
for fields $Q_i$ without large Yukawa couplings, the dominant contributions 
to $\dot{\gamma}_i$ are proportional to the beta functions of gauge couplings.
Since the $SU(2)$ and $U(1)$ gauge groups of electroweak theory are both 
asymptotically nonfree in the MSSM, the color-blind sleptons receive large
negative contributions to their squared masses.  The minimum of the potential 
in the above theory breaks electromagnetism near the scale of the soft masses
ruling out the model.

One otherwise remarkable feature of the spectrum which compounds the
slepton mass problem is its UV insensitivity.  Though the soft
parameters are generated at the Planck scale, the functional form of
Eqs.~(\ref{eq:A}, \ref{eq:m}, \ref{eq:gaugino}) are valid at the weak
scale.  The fact that they indeed solve the RGEs at all orders 
in perturbation theory provides a non-trivial check.  
No matter how complicated physics is at high energies, 
which presumably distinguish flavor in order to generate fermion 
mass hierarchies via, {\it e.g.}\/, the Froggatt--Nielsen mechanism 
\cite{Froggatt:1979nt}, the low-energy soft supersymmetry-breaking 
parameters are uniquely determined by a single parameter 
$m_{3/2}$ and hence the flavor breaking is only given by Yukawa couplings.  
This makes the constraints from the lack of flavor-changing processes 
essentially automatically satisfied.

Note that the term UV insensitivity refers to the lack of dependences
on the particles and interactions in the ultraviolet in the observable
sector.  The way the supersymmetry breaking effects are induced in the
observable sector definitely depends on the coupling between the
hidden and observable sectors.  The point is that once a particular
coupling between two sectors is specified, in this case by putting
them on different branes with no light moduli mediating supersymmetry
breaking in the bulk \cite{Randall:1999uk}, we obtain supersymmetry
breaking effects reviewed in this section no matter what particles and
interactions there are in the observable sector.  This is the property
preserved in the modification of the anomaly-mediated supersymmetry
breaking which will be discussed in the next section.

\section{$D$-term Contributions\label{sec:D}}

In the previous section, we reviewed how the supersymmetry breaking in 
the Weyl compensator field $\phi = 1 + \theta^{2} m_{3/2}$ induces 
supersymmetry-breaking effects via superconformal anomalies.  Here we 
extend the discussion by including a fictitious abelian gauge field 
$V = \theta^{2} \bar{\theta}^{2} D$ with $D\neq 0$.  We show that the 
low-energy supersymmetry-breaking parameters are completely 
insensitive to physics in the UV analogous to the purely 
anomaly-mediated case.
The RGE property under the $D$-term background was also discussed in 
\cite{Jack:2000cd,Hisano:1999tm}.  Here we give an all-order proof 
using superfield spurion formalism and also discuss the complete UV 
insensitivity.  We use the notations of \cite{Arkani-Hamed:1998kj}.  

The idea is to identify an anomaly-free $U(1)$ symmetry in the 
theory and to pretend that it is gauged.  This introduces a 
fictitious gauge multiplet without a kinetic term.  We assume that the 
gauge multiplet has a supersymmetry-breaking spurion in its 
$D$-component:
\begin{equation}
        V = \theta^{2} \bar{\theta}^{2} D.
\end{equation}
If needed, we can further introduce a ``spectator'' field to cancel 
the $U(1)^{3}$ anomaly as well, which has no interactions with 
physical fields in the superpotential.  But because this gauge 
multiplet is non-dynamical, the spectator fields then do not have any 
interactions from the rest, and hence do not modify the discussions 
at all except for justifying the ``gauging'' of the $U(1)$ 
symmetry fictitiously.  

Because we know the $U(1)$ charges of the matter fields under this 
fictitious $U(1)$ gauge symmetry, it uniquely determines how the 
$D$-term couples to the matter fields, and the coefficients 
${\cal Z}_{i}$ are changed to
\begin{equation}
        {\cal Z}_{i} \rightarrow {\cal Z}_{i} e^{q_{i} V},
        \label{eq:addD}
\end{equation}
where $q_{i}$ is the $U(1)$ charge of the chiral multiplet $Q_{i}$.  This 
coupling contributes to the scalar masses
\begin{equation}
        m^{2}_{i} \rightarrow \bar{m}^{2}_{i} = m^{2}_{i} - q_{i} D.
\label{eq:traj}
\end{equation}
It is easy to see that if $m^{2}_{i}(\mu)$ is a RGE trajectory, so 
is $\bar{m}^{2}_{i}(\mu)$.  The point is that the counter term to
the kinetic term takes the form ${\cal Z}_{0} = {\cal Z}(\mu) + \delta 
{\cal Z}$, where
\begin{equation}
        \delta {\cal Z}_{i} = {\cal Z}_{i} C\left( 
        \frac{|\lambda_{jkl}|^{2}}{{\cal Z}_{j} {\cal Z}_{k} {\cal Z}_{l}},
        \frac{|\Lambda|}{\mu},
        S + S^{\dagger} - \frac{1}{4\pi^{2}} \sum_{j} T_{j} \ln {\cal Z}_{j}
        \right),
\end{equation}
because of the reparameterization invariances
\begin{eqnarray}
        Q_{i} \rightarrow e^{A_{i}} Q_{i}, &\qquad&
        {\cal Z}_{i} \rightarrow e^{-(A_{i}+A_{i}^{\dagger})} {\cal Z}_{i}, 
        \nonumber \\
        S \rightarrow S - \frac{1}{4\pi^{2}} \sum_{j} T_{j} A_{j}, &\qquad&
        \lambda_{ijk} \rightarrow e^{-(A_{i}+A_{j}+A_{k})} \lambda_{ijk}.
\end{eqnarray}
The shift in $S$ is due to the Konishi anomaly \cite{Konishi,AM}.  In
the function $C$, the dependence on the superpotential couplings have
the product of three kinetic coefficients ${\cal Z}_{j} {\cal Z}_{k}
{\cal Z}_{l}$.  Under the modification Eq.~(\ref{eq:addD}), this
product changes by $e^{(q_{j}+q_{k}+q_{l})V}$.  However, the
fictitious $U(1)$ gauge invariance demands that $q_{j}+q_{k}+q_{l} =
0$ for the superpotential coupling $\lambda_{jkl} Q_{j} Q_{k} Q_{l}$
to be allowed, and hence there is no dependence on $V$ here.  Another
possible dependence is in $S + S^{\dagger} - \frac{1}{4\pi^{2}}
\sum_{j} T_{j} \ln {\cal Z}_{j}$, where the possible shift is given by
$- \frac{1}{4\pi^{2}} \sum_{j} T_{j} q_{j} V$.  However, the condition
that the fictitious $U(1)$ symmetry is anomaly-free under the true
gauge groups $\sum_{j} q_{j} T_{j} = 0$ tells us that this possible
shift actually vanishes.  Thus, an RGE trajectory remains an RGE
trajectory even after the change Eq.~(\ref{eq:addD}).  
Therefore, the anomaly-mediated supersymmetry-breaking parameters,
together with the $D$-term contributions, would solve the RGE and are
hence consistent.

The main result of this paper is that the anomaly-mediated 
supersymmetry breaking together with the $D$-term contributions 
preserves the UV insensitivity of the purely anomaly-mediated case.  
The argument is very simple.  No matter what the high-energy theory 
may be, it should preserve the fictitious $U(1)$ gauge symmetry whose 
gauge multiplet contains the $D$-term background.  After integrating 
out the heavy fields, the low-energy fields should still satisfy the 
same fictitious $U(1)$ gauge symmetry, which uniquely determines the 
$D$-term contributions.  The necessary and sufficient condition 
for this to hold is that the fictitious $U(1)$ gauge symmetry is not 
spontaneously broken by high energy physics.

It is important that there is an additional operator in the low-energy 
theory consistent with the fictitious $U(1)$ gauge invariance if there 
are $U(1)$ gauge fields, such as $U(1)_{Y}$ in the standard model. From 
the fictitious $e^{V}$, one can construct the fictitious 
$W_{\alpha} = \theta_{\alpha} D$.  Then the operator
\begin{equation}
  c \int d^{2} \theta\, W_{\alpha} W_{Y}^{\alpha} = c D D_{Y}
  = \xi_Y D_Y,
\end{equation}
is allowed, and this is nothing but the Fayet--Iliopoulos $D$-term
$\xi_Y \equiv c D$ for the $U(1)_{Y}$ gauge multiplet.  $c$ is an
arbitrary coefficient for this operator.  Therefore, in the low-energy
theory, all the scalar masses-squared are shifted by $q_{i} D$, and
there is a possible Fayet--Iliopoulos $D$-term for each of the $U(1)$
gauge factors, with arbitrary coefficients.

Note that the spurion parameter $D$ does not run, while $\xi_Y$ does
being a renormalizable coupling.  For example, $\xi_Y$ receives
renormalization proportional to ${\rm Tr}\,q_i Y_i$ at the one-loop
level.  An interesting point to check is the reparameterization
invariance of the $D$-terms.  Since the total masses-squared including
the Fayet--Iliopoulos term is
\begin{equation}
        m^{2}_{i}(\mu) \rightarrow \bar{m}^{2}_{i}(\mu) =
        m^{2}_{i}(\mu) - q_{i} D + g_Y^2(\mu) Y_i \xi_Y(\mu),
\end{equation}
one can change the definition of the $U(1)$ charge $q_i$ as $q'_i =
q_i + \eta Y_i$ while changing the Fayet--Iliopoulos term as $\xi'_Y =
\xi_Y + \eta D/g_Y^2$.  The renormalization of $\xi'_Y$ also changes
accordingly including now a term proportional to ${\rm Tr}\, q'_i Y_i$.
The running of soft parameters is unaffected iff $D$ is unrenormalized
in both bases because the renormalization of $\xi'_Y$ changes and
cancels the redefinition of other parameters.

\section{Model\label{sec:model}}

It is easy to justify the assumptions of our picture, 
namely the controlled introduction of supersymmetry breaking through 
the Weyl compensator and the $D$-term of a fictitious gauge multiplet.  

It was pointed out in \cite{Randall:1999uk} that a physical 
separation between the observable and the hidden sectors by putting 
them on different branes results in their separation in the K\"ahler 
density, and hence the K\"ahler potential takes the form
\begin{equation}
        K = -3 \ln \left( 1 - \frac{1}{3} f_{\rm obs}(Q,Q^{\dagger})
                - \frac{1}{3} f_{\rm hid}(H,H^{\dagger}) \right),
\end{equation}
where $H$ denotes the hidden sector fields and $f_{\rm obs}$, 
$f_{\rm hid}$ are arbitrary real functions.  
The superpotential has also a similar separation 
between observable and hidden pieces.  This separation was shown to 
give vanishing soft masses-squared at the tree-level and the 
supersymmetry breaking is given purely by anomalies.

We propose an additional $U(1)$ gauge field in the bulk, 
which is broken on the hidden brane.  The anomaly cancellation 
on the observable brane is obviously required.  Because the K\"ahler 
potential has two independent functions $f_{\rm obs}$ and $f_{\rm hid}$, 
even though the true gauge invariance is only a single $U(1)$, 
there are two {\it global}\/ $U(1)$ symmetries acting on 
observable and hidden fields separately.  
This $U(1) \times U(1)$ is reduced to the diagonal subgroup only 
through charged bulk fields coupling to both branes.
If these charged fields are heavy, therefore, the two global $U(1)$'s 
are preserved up to exponentially small effects.
After the $U(1)$ group is broken on the hidden brane, the 
zero-mode gauge multiplet can pick up a non-vanishing $D$-component 
due to the supersymmetry-breaking effects on the hidden brane.  This 
effect introduces $e^{V}$ with $V = \theta^{2} \bar{\theta}^{2} D$ 
as discussed in the previous sections.  

A simple toy model is given as follows.
Let us consider three chiral superfields $X$, $\psi$ and $\bar{\psi}$ 
living on the hidden brane with the $U(1)$ charges $X(0)$, 
$\psi(+1)$ and $\bar{\psi}(-1)$.
We introduce the superpotential
\begin{equation}
  W = X (\psi \bar{\psi} - \mu^2),
\end{equation}
which forces $\psi$, $\bar{\psi}$ to acquire expectation values.
Now, since $\psi$ and $\bar{\psi}$ fields are living on the 
hidden brane, they can directly couple to the field $z$ which breaks
supersymmetry by its $F$-component in the K\"ahler potential, 
{\it i.e.}\/, $K \ni \frac{c}{M_*^2} z^{\dagger} z \psi^{\dagger} \psi 
+ \frac{\bar{c}}{M_*^2} z^{\dagger} z \bar{\psi}^{\dagger} \bar{\psi}$.
These couplings generate soft masses for the $\psi$ and $\bar{\psi}$ 
fields in addition to the anomaly-mediated ones.
If $c$ and $\bar{c}$ are different, then, the two soft masses are 
different, $m^2_\psi \neq m^2_{\bar{\psi}}$, resulting in different 
expectation values for the $\psi$ and $\bar{\psi}$ fields.
This causes nonvanishing $D$-term expectation value for the 
gauge multiplet at low-energy, giving soft masses for the observable 
fields proportional to their $U(1)$ charges.
Since the $D$-term expectation value is of order 
$m_\psi \simeq c F_z / M_*$, however, we need somewhat small coefficients 
$c \sim \bar{c} \sim 1/((16 \pi^2)^2 M_* R)$ to make the $D$-term 
contribution comparable to the anomaly mediated one.
Here, $R$ is the compactification radius for the extra dimension.

The above dynamics can also be understood in the following way 
using superfield spurion language.
The K\"aher potential for the $\psi$ and $\bar{\psi}$ fields 
can be written as
\begin{equation}
  K = \psi^{\dagger} e^{U+V} \psi + 
      \bar{\psi}^{\dagger} e^{\bar{U}-V} \bar{\psi},
\end{equation}
where $V$ is the $U(1)$ gauge multiplet, while 
$U = - \theta^2 \bar{\theta}^2 m_{\psi}^2$ and 
$\bar{U} = - \theta^2 \bar{\theta}^2 m_{\bar{\psi}}^2$ 
parameterize the soft masses for the $\psi$ and $\bar{\psi}$ fields 
coming from direct couplings to the $z$ field.
Now we integrate out all the heavy fields in the theory 
in a supersymmetric manner.
When $\psi$ and $\bar{\psi}$ acquire expectation values, we can 
go the unitary gauge $\psi = \bar{\psi} = \mu$ and find
\begin{equation}
  K = \mu^2 (e^{U+V} + e^{\bar{U}-V}).
\end{equation}
Minimizing it with respect to $V$, we find $V = -(U-\bar{U})/2 = 
\theta^2 \bar{\theta}^2 (m_\psi^2 - m_{\bar{\psi}}^2)/2$, and hence
the gauge multiplet acquires an expectation value in its
$D$-component.  Since the gauge multiplet couples to matter fields $Q_i$ 
on the observable brane such as
\begin{equation}
  K = Q_i^{\dagger} e^{q_i V} Q_i,
\end{equation}
it induces $D$-term contribution to the matter scalar masses according
to their $U(1)$ charges $q_i$.

An important point is that the {\it global}\/ $U(1)$ symmetry in 
the effective theory of the observable brane is not broken even after 
$\psi$ and $\bar{\psi}$ fields have expectation values, because of 
the accidental global $U(1)$ symmetry in the K\"ahler
potential at the tree-level.  
Therefore, the interactions among the observable fields respect 
the $U(1)$ invariance which is necessary for the soft masses to be on the 
trajectory defined by Eq.~(\ref{eq:traj}).
One may worry that loop diagrams by the
bulk $U(1)$ gauge field can break the $U(1)$ symmetry on the
observable brane; however such effects can be made arbitrarily small
by making the $U(1)$ gauge coupling constant small, while the
$D$-term contribution to the scalar masses is independent of the 
size of the $U(1)$ gauge coupling constant, as we have seen already 
explicitly and has been known for some time \cite{KMY}.

One can also make a more realistic model in which the 
anomaly-mediated and $D$-term contributions appear naturally 
at the same order.
Let us consider that the $U(1)$-breaking fields $\psi$ and 
$\bar{\psi}$ live on the third brane which is different from 
the observable and the supersymmetry-breaking branes.
Suppose that $\psi = \bar{\psi}$ direction is a flat direction, 
and $\psi$ and $\bar{\psi}$ have different interactions in the 
superpotential.  Then, the anomaly-mediated contribution generates 
expectation values for these fields but two expectation values are 
slightly different, so that there remains nonvanishing $D$-term 
of the order of the soft masses generated by anomaly mediation.
This $D$-term contribution is nondecoupling and gives soft 
supersymmetry-breaking masses to the observable fields proportional 
to their $U(1)$ charges \cite{Pomarol:1999ie}.
A crucial difference from the previous toy model is that the resulting 
$D$-term contribution is automatically the same order with 
the anomaly-mediated one.
Once again, an important point here is that since $\psi$ and 
$\bar{\psi}$ fields live on a different brane than the observable one, 
they do not directly couple to the standard-model fields.  
Therefore, no matter how large the $U(1)$ gauge symmetry is broken
by the expectation values of the $\psi$ and $\bar{\psi}$ fields, 
the interactions among the standard-model fields still respect the 
$U(1)$ invariance.
The only assumption needed is that there is no light bulk field 
which carries nonvanishing $U(1)$ charge.

In order to understand the importance of breaking the bulk $U(1)$ 
on a different brane, consider the following simple toy model 
where direct couplings between the $U(1)$ breaking fields and 
the standard-model fields are allowed. 
In addition to the light fields, there are superheavy fields 
$\Psi, \bar{\Psi}$ with $U(1)$ charges of $-2, +2$ and a field $\varphi$ 
with charge $+1$ which gets an expectation value to break the 
$U(1)$ symmetry. Suppose a light field $l$ also has charge $+1$. 
Then we can write the relevant interactions
\begin{equation}
  W = l \Psi \varphi + M_\Psi \Psi \bar{\Psi}.
\end{equation}
Upon integrating out $\Psi$ and $\bar{\Psi}$, we generate the 
following term in the K\"ahler potential:
\begin{equation}
  \frac{1}{M_\Psi^2}(\varphi l)^{\dagger} e^{2 V} (\varphi l).
\label{eq:dang-op}
\end{equation}
Once $\varphi$ acquires its expectation value, and inserting the
$D$-term for $V$, this gives a contribution to the $l$ soft mass which
is not proportional to its $U(1)$ charge and is off the trajectory. Of
course, $M_\Psi$ preserves $U(1)$ and hence there is no reason for it
to be as low as $\varphi$; if $\varphi \ll M_\Psi$ the deviation from
the trajectory can be suppressed. But $\varphi \ll M_\Psi$ is an
additional assumption spoiling the spirit of UV insensitivity we are
seeking.

Note that the $U(1)$ hypercharge gauge field is localized on the 
observable brane and does not get a $D$-term in the above way.
However, once the bulk $U(1)$ gauge multiplet obtains the $D$-term 
expectation value, the Fayet--Iliopoulos $D$-term for other $U(1)$ 
gauge multiplets on the observable brane can be generated either from 
the tree-level kinetic mixing terms $\int d^{2} \theta\, 
W_{i\alpha} W_{j}^{\alpha}$ or from loop corrections.  Specifically, 
the Fayet--Iliopoulos term for the $U(1)$ hypercharge can be generated 
in these ways.

Therefore, the features necessary to introduce supersymmetry breaking 
in the controlled way we desire, through the $F$-component of the Weyl 
compensator field as well as the $D$-term of a spurious background 
gauge field, can be naturally obtained by separating the observable 
and hidden sectors on different branes with a bulk $U(1)$ gauge 
field broken on a brane other than the observable one.

\section{The (N)MSSM\label{sec:NMSSM}}

In the MSSM, the squarks and sleptons have known gauge quantum numbers 
and Yukawa couplings and hence their masses from anomaly mediation 
are completely calculable.  This high degree of predictivity results 
in disastrous negative masses-squared for the sleptons and hence the 
model is excluded.  However, as shown in the previous section, we can 
introduce $D$-terms to modify the scalar masses-squared and it is 
interesting to ask if such a modification can make all slepton 
masses-squared positive.  

The first question is what non-anomalous $U(1)$ symmetries are there
in the MSSM. Assuming there are also neutrino masses (Dirac masses for
the sake of the discussions here) and mixing among the leptons, there
are only two non-anomalous $U(1)$ symmetries: $U(1)_{Y}$ and
$U(1)_{B-L}$.  Therefore we can introduce two additional parameters to
the anomaly mediated supersymmetry breaking, $D_{Y}$ and $D_{B-L}$.
Here and below, we use the notation $D_Y = - g_Y^2 \xi_Y$.  In the
framework of higher dimensional models discussed in the previous
section, the $U(1)_{B-L}$ gauge field is identified with the bulk
$U(1)$ gauge field and $D_Y$ is generated through kinetic mixing 
between $U(1)_{B-L}$ and $U(1)_{Y}$.  The slepton masses are then 
modified as
\begin{eqnarray}
        m^{2}_{\tilde{L}} &\rightarrow& \bar{m}^{2}_{\tilde{L}} 
        = m^{2}_{\tilde{L}} + \frac{1}{2} D_{Y} + D_{B-L}, \\
        m^{2}_{\tilde{e}} &\rightarrow& \bar{m}^{2}_{\tilde{e}}
        = m^{2}_{\tilde{e}} - D_{Y} - D_{B-L},
\end{eqnarray}
where $m^{2}_{\tilde{f}}$ represents the pure anomaly-mediated
contribution.  A necessary condition for a viable spectrum is
\begin{equation}
D_Y < -D_{B-L} < {1\over 2} D_Y < 0.
\end{equation}
It has been shown that the spectrum can in fact be made viable in some
region of parameter space \cite{Jack:2000cd}.

As for the particles in the supersymmetric standard model, we find
the gaugino masses
\begin{eqnarray}
  M_1 &=& 1.43 M, \nonumber \\
  M_2 &=& 0.414 M, \\
  M_3 &=& -3.67 M, \nonumber
\end{eqnarray}
where $M = m_{3/2}/(16\pi^2)$.  The masses for the first and second
generation scalars with Yukawa couplings neglected are
\begin{eqnarray}
  m^2_{\tilde{L}} &=& -0.351 M^2 + \frac{1}{2} D_{Y} + D_{B-L},
\nonumber \\
  m^2_{\tilde{e}} &=& -0.374 M^2 - D_{Y} - D_{B-L},
\nonumber \\
  m^2_{\tilde{Q}} &=& 11.7 M^2 - \frac{1}{6} D_Y - \frac{1}{3} D_{B-L},
\\
  m^2_{\tilde{u}} &=& 11.8 M^2 + \frac{2}{3} D_Y + \frac{1}{3} D_{B-L},
\nonumber \\
  m^2_{\tilde{d}} &=& 11.9 M^2 - \frac{1}{3} D_Y + \frac{1}{3} D_{B-L},
\nonumber
\end{eqnarray}
while for the third generation scalars and Higgs bosons
\begin{eqnarray}
  m^2_{\tilde{L}_3} &=& -0.352 M^2 + \frac{1}{2} D_{Y} + D_{B-L},
\nonumber \\
  m^2_{\tilde{e}_3} &=& -0.377 M^2 - D_{Y} - D_{B-L},
\nonumber \\
  m^2_{\tilde{Q}_3} &=& 9.55 M^2 - \frac{1}{6} D_Y - \frac{1}{3} D_{B-L},
\nonumber \\
  m^2_{\tilde{u}_3} &=& 7.54 M^2 + \frac{2}{3} D_Y + \frac{1}{3} D_{B-L},
\\
  m^2_{\tilde{d}_3} &=& 11.9 M^2 - \frac{1}{3} D_Y + \frac{1}{3} D_{B-L},
\nonumber \\
  m^2_{H_u}       &=& -6.72  M^2 - \frac{1}{2} D_Y,
\nonumber \\
  m^2_{H_d}       &=& -0.402 M^2 + \frac{1}{2} D_Y,
\nonumber
\end{eqnarray}
for $\tan\beta = 3$ as an example.\footnote{
These mass parameters have been evaluated at a scale $\mu = 500$ GeV, 
for simplicity.  To obtain the physical masses, superparticle threshold 
corrections should be taken into account appropriately.}

An interesting feature of the above spectrum is that while the scalar 
masses receive contributions from $D$-terms, the gaugino masses remain
those of anomaly mediation.
Thus, in a broad region of the parameter space, the lightest 
supersymmetric particle is the neutral wino nearly degenerate with 
the charged one, giving rise to signatures discussed in \cite{signal}.
But the sleptons can also be the lightest supersymmetric particle 
depending on the parameters $M, D_Y$ and $D_{B-L}$.
Furthermore, since the scalar masses are given by only three parameters, 
there are a number of sum rules for the scalar masses as 
derived in \cite{Jack:2000cd}.

Note that the right-handed neutrino superfield $N$, which in this 
minimal model only picks up soft masses from the $D$-term and the 
(miniscule) anomaly-mediated contribution from the tiny neutrino 
Yukawa coupling, gets a {\it negative} mass squared. However, this 
difficulty is easily remedied by e.g. adding additional couplings 
to new standard-model singlets charged under $U(1)_{B-L}$. 
For instance, consider new fields $X,\bar{X}$ with charges $-2, +2$ and 
a singlet $Y$ with zero charge under $U(1)_{B-L}$. Now add the couplings 
$Y X \bar{X} + N^2 X$. Then, the scalar components of $N,Y,X,\bar{X}$ 
can pick up net positive soft masses from anomaly and $D$-term 
mediations.

Now it is well-known that the MSSM probably does not work within the
anomaly-mediated supersymmetry breaking because the $\mu$ term is
associated with a too-large $B\mu$ term because $B \sim m_{3/2}$
rather than $m_{3/2}/(16\pi^2)$.  The next simplest possibility is the
NMSSM with the superpotential
\begin{equation}
  \lambda S H_u H_d + h S^3,
\end{equation}
where $S$ is a new standard model singlet chiral superfield.  Because
$m^2_{H_u}$ is negative and $m^2_{H_d}$ less so with the above
spectrum, it is reasonable to expect that one can find a correct
electroweak symmetry breaking.  The details of the phenomenological
analysis is beyond the scope of this paper and will be discussed
elsewhere.  

\section{Neutrino Masses\label{sec:neutrino}}

Since we have introduced a fictitious $U(1)_{B-L}$ gauge symmetry, a
natural question is how we can explain small neutrino masses.  As
mentioned earlier, Dirac neutrinos with small Yukawa couplings,
possibly as consequences of flavor symmetries, is definitely a natural
possibility within our framework.  However, one can resort to other
mechanisms to explain small neutrino masses consistent with our
framework as well.

For instance, we can consider the situation where the K\"ahler potential 
contains a term $\frac{1}{M_*} L H_u N + {\rm h.c.}$ while the 
superpotential coupling $LH_u N$ is absent.  This possibility is 
naturally realized using a global $U(1)_R$ symmetry, for example.  
Then, the K\"ahler potential term picks up the Weyl compensator
\begin{equation}
  \frac{\phi^\dagger}{\phi^2} \frac{1}{M_*} L H_u N,
\end{equation}
and by expanding $\phi^\dagger = 1 + \bar{\theta}^2 m_{3/2}$, we find
a term which is effectively a superpotential $LH_uN$ with a coupling
constant of the order of $m_{3/2}/M_* \sim 16\pi^2 v (M_{\rm Planck}
R)^{1/3}/M_{\rm Planck}$ \cite{AHMSW,BN}.  This size for the Dirac 
neutrino Yukawa coupling is remarkably consistent with the size 
expected from the atmospheric neutrino data.

Another possibility is to write down the mass term for the
right-handed neutrino by hand: $\frac{M_R}{2} N N$ in the
superpotential, giving the standard seesaw mechanism \cite{Seesaw}.  
This of course breaks $U(1)_{B-L}$ explicitly, but only softly, 
and does not change the RGEs of soft supersymmetry-breaking 
parameters at all.  The only concern is the threshold corrections 
when the right-handed neutrinos are decoupled.  
It is interesting that the threshold corrections are
still exactly the same with supersymmetric thresholds which keep the
supersymmetry-breaking parameters on the trajectories of our framework
at the leading order.  One may imagine that such $U(1)_{B-L}$-breaking 
mass parameter was induced on our brane by a field carrying $U(1)_{B-L}$
charge in the bulk, or even simply by a spontaneous breakdown occurring 
on our brane since dangerous operators such as those given in 
Eq.~(\ref{eq:dang-op}) are generically generated only at the loop level.
In this framework, however, the supersymmetry-breaking 
parameters are not exactly on the trajectories below the scale of 
$U(1)_{B-L}$ breaking, and one expects corrections at higher orders 
in $1/(16\pi^2)$.

\section{Conclusion\label{sec:conclusion}}

Anomaly mediation of supersymmetry breaking is attractive because
the soft supersymmetry breaking parameters at a given energy scale are
determined only by physics at that energy scale (UV insensitivity) and
hence is highly predictive (only one parameter).  The resulting
supersymmetry breaking parameters are automatically safe from
generating too-large flavor-changing neutral current effects no matter
how complicated flavor physics at high-energy scales is.  However the
anomaly mediation failed phenomenologically because of its high
predictivity: slepton masses-squared are negative.  It is known that
one can add $D$-term contributions for $U(1)_Y$ and $U(1)_{B-L}$ to
the anomaly-mediated supersymmetry breaking to make the superparticle
spectrum phenomenologically viable.  In this paper we have shown that
one can indeed add these $D$-terms keeping the complete UV
insensitivity no matter how complicated thresholds there are.  The
only assumption is that the $U(1)_{B-L}$ is a global symmetry in the
supersymmetric standard model.

This framework can be derived from supersymmetry breaking and
$U(1)_{B-L}$ breaking on hidden brane(s).  The $U(1)_{B-L}$ gauge
multiplet lives in the bulk but it is broken at a high scale unlike in
the gaugino mediation.  Assuming no light bulk fields charged under 
$U(1)_{B-L}$ (similar to the usual assumption in anomaly mediation), 
the physics on the observable brane has a global $U(1)_{B-L}$ symmetry 
up to exponentially suppressed effects.  The supersymmetry 
breaking effects are described in terms of the Weyl compensator 
with an $F$-component expectation value, a non-dynamical $U(1)_{B-L}$ 
gauge multiplet with a $D$-component expectation value, and
a Fayet--Iliopoulos term for $U(1)_Y$.  Because the low-energy
supersymmetry breaking is constrained by the superconformal and
$U(1)_{B-L}$ invariances, the soft parameters are determined
completely by the physics at the energy scale of interest and hence UV
insensitive.

Despite the $U(1)_{B-L}$ invariance, one can still generate neutrino
masses at the correct energy scale for atmospheric neutrino
oscillations naturally due to an operator at the fundamental scale.

\section*{Acknowledgements}

Y.N. thanks the Miller Institute for Basic Research in Science 
for financial support.
This work was supported in part by the U.S. Department of Energy 
under Contracts DE-AC03-76SF00098, DE-FG02-90ER40560 and W-31-109-ENG-38,
and in part by the National Science Foundation under grant PHY-95-14797.

\end{document}